\documentclass[letterpaper,twocolumn,10pt]{article}\usepackage{usenix}

\usepackage{tikz}
\usepackage{amsmath}

\usepackage{tabularx}
\usepackage{algorithm}
\usepackage{algpseudocode}

\usepackage{amsmath,amsfonts}
\usepackage{graphicx}
\usepackage{textcomp}
\usepackage{xcolor}
\usepackage{booktabs}

\usepackage{url}
\usepackage{comment}
\usepackage{bm}
\usepackage{hyperref}
\usepackage{todonotes}
\usepackage{csquotes}
\usepackage{enumitem}

\usepackage{xcolor}
\usepackage[font=small,]{caption}
\usepackage{etoolbox}

\usepackage{xspace}

\usepackage[skip=0.5pt]{caption} 

\usepackage{graphicx}
\usepackage{subcaption}

\newcommand{\defenseheader}{\texttt{HTTP-Sync}\xspace}
\newcommand{\defenseheaderhmac}{\texttt{HTTP-Sync-HMAC}\xspace}

\usepackage{listings}
\usepackage{xcolor}
\usepackage{caption}
\usepackage{array} 
\usepackage{tabularx}

\definecolor{verbBlue}{RGB}{0,0,255}        
\definecolor{hdrLightGreen}{RGB}{94,109,30} 
\definecolor{httpGreen}{RGB}{0,116,0}       
\definecolor{textGray}{gray}{0.1}           
\definecolor{frameGray}{gray}{0.6}          
\definecolor{bgGray}{RGB}{250,250,250}      

\lstdefinestyle{httpcode}{
  basicstyle=\ttfamily\small\color{textGray},
  frame=tb,
  lineskip=-6.1875076pt,       
  xleftmargin=0pt,
  xrightmargin=0pt,
  keepspaces=true,
  columns=fullflexible,
  showstringspaces=false,
  breaklines=false,
  upquote=true,
  escapeinside={(*@}{@*)}, 
  showtabs=false,
  showspaces=false,
  showlines=true,       
  keepspaces=true,      
  literate={\\}{{\textbackslash}}1
           {\{}{{\{}}1
           {\}}{{\}}}1
           {\~}{{\textasciitilde}}1,
  moredelim=**[is][\color{verbBlue}]{||}{||},           
  moredelim=**[is][\color{verbBlue}\bfseries]{<<}{>>},  
  moredelim=**[is][\color{hdrLightGreen}]{[[}{]]}, 
  moredelim=**[is][\color{httpGreen}\bfseries]{((}{))},     
}

\DeclareCaptionType{listing}[Listing][List of Listings]

\begin{document}

\date{}

\title{\Large \bf HTTP Request Synchronization Defeats Discrepancy Attacks}

\author{
{\rm Cem Topcuoglu$^{*}$, Kaan Onarlioglu$^{*\dagger}$, Steven Sprecher$^{*}$, Engin Kirda$^{*}$}\\
$^{*}$Northeastern University, $^{\dagger}$Akamai Technologies\\
}

\makeatletter
\AtBeginEnvironment{noerr}{\dontdofcolorbox}
\def\dontdofcolorbox{\renewcommand\fcolorbox[4][]{##4}}
\makeatother
\newenvironment{noerr}{}

\newcommand{\mut}[2]{{\color{#1}#2}}
\newcommand{\nginxserver}{NGINX\xspace}
\newcommand{\envoyserver}{Envoy Proxy\xspace}
\newcommand{\fastlyserver}{Fastly\xspace}

\newcommand{\twopartdef}[4]
{
	\left\{
		\begin{array}{ll}
			#1 & \mbox{if } #2 \\
			#3 & \mbox{if } #4
		\end{array}
	\right.
}

\maketitle

\begin{abstract}
Contemporary web application architectures involve many layers of proxy services that process traffic. Due to the complexity of HTTP and vendor design decisions, these proxies sometimes process a given request in different ways. Attackers can exploit these processing discrepancies to launch damaging attacks including web cache poisoning and request smuggling. Discrepancy attacks are surging, yet, there exists no systemic defense.

In this work, we propose the first comprehensive defense to address this problem, called HTTP Request Synchronization. Our scheme uses standard HTTP extension mechanisms to augment each request with a complete processing history. It propagates this context through the traffic path detailing how each server hop has processed said request. Using this history, every proxy server can validate that their processing is consistent with all previous hops, eliminating discrepancy attacks. We implement our scheme for 5 popular proxy technologies, Apache, NGINX, HAProxy, Varnish, and Cloudflare, demonstrating its practical impact.
\end{abstract}

\section{Introduction}

Contemporary web application architectures rely on many layers of HTTP proxies serving as load balancers, caches, web application firewalls, and similar services that filter, transform, and route traffic. Each of these proxies are HTTP servers at their core, parsing and processing requests according to the HTTP protocol specification.

Sadly, the HTTP specification is decidedly complex and ambiguous in parts, requiring server developers to fill in the blanks with \textit{reasonable} implementation choices. Even assuming that the resulting implementation is free of bugs, these design decisions often result in servers from different vendors processing identical HTTP requests inconsistently. That, in turn, leads to surprising interactions in proxied web applications where servers on the traffic path disagree on how a given request should be processed.

Attacks exploiting such processing discrepancies, also known as \textit{discrepancy attacks}, have been documented for almost two decades (e.g.,~\cite{smuggling_old}), yet have remained unattended by the security community. However, the proliferation of complex proxied cloud architectures and Content Delivery Networks (CDNs) have led to a surge of a wide variety of discrepancy attacks such as cache poisoning, request smuggling, path confusion, and host name confusion~\cite{host-of-troubles,cache_poisoning,smuggling_new,sec2022frameshifter,jabiyev2021t}.

While there have been one-off solutions proposed against specific instances of discrepancy attacks (e.g., Cloudflare's Cache Deception Armor~\cite{wcd_armor}, Amazon's HTTP Desync Guardian~\cite{aws_guardian}), the security industry has mostly been reactive to new discoveries by improving Web Application Firewall rules to detect known payloads, or patching servers to eliminate problematic behavior~\cite{cloudflare_cpdos, varnish_cve, ats_cve, nginx_cve, haproxy_patched, akamai_kaan, akamai_waf}. Likewise, academia has focused their efforts on measurement and automated attack discovery~\cite{jabiyev2021t, jabiyev2024gudifu,hdiff,cpdos,mirheidari2020cached,mirheidari2022web}, but not defense. To date, there exists no scalable, general defense against discrepancy attacks.   

Here, we present the first general design to secure proxied web applications against discrepancy attacks. As opposed to the aforementioned attempts at eliminating each implementation quirk that leads to a vulnerability, our solution addresses the underlying issue. We do this by providing a mechanism for every server on the traffic path to validate whether their request processing is consistent with previous hops. Specifically, we use standard HTTP extension mechanisms to propagate information on parsed sensitive request components, compiling an unforgeable request processing history that can be validated by any server on the path. Any discrepancy in this history implies a potential attack or unintended safety issue, signaling to the server to terminate the connection. We call this scheme \textit{HTTP Request Synchronization.}

We implement our general design focusing on the specific request components targeted by known attacks. Namely, the full URL, the \texttt{Host} header value, and the request message framing (i.e., the \texttt{Content-Length} header value, or individual chunk lengths for chunked body encoding indicated by the \texttt{Transfer-Encoding: chunked} header). Our implementation can be trivially extended to cover any other request component as future developments necessitate. We supplement HTTP requests with this new integrity information only via mechanisms already provided in the standard, without requiring any changes to the existing protocol specification. We are fully compliant with HTTP RFCs.

We provide and evaluate prototype implementations for 4 popular server technologies (Apache, NGINX, HAProxy, Varnish), demonstrating that the design is achievable with practical modifications to the underlying technology with promising performance. We also provide a prototype implementation for a popular CDN, Cloudflare, using their edge compute technology Cloudflare Workers, demonstrating an application to cloud platforms with no source code modification necessary. Our solution also works correctly in the presence of oblivious servers that are unaware of our work.

Finally, we empirically show the effectiveness of our design through case studies based on real-life attacks.
 
To reiterate, we make the following contributions:
\begin{itemize}
    \item We present \textit{HTTP Request Synchronization}, the first general defense against discrepancy attacks.
    \item We implement our defense on 5 popular proxy technologies, demonstrating that the design is practical.
    \item We empirically show the effectiveness of our defense over 3 case studies drawn from real processing discrepancy attacks.
\end{itemize}
\section{Background and Related Work}
\label{section:background}


The discussion and examples in the rest of this paper are based on HTTP version 1.1, which is not a limitation or omission but intentional, because HTTP/1.1 is the most susceptible to discrepancy attacks.

\subsection{HTTP Essentials}

Contemporary web application architectures often consist of multiple layers of reverse proxies deployed either as stand alone boxes, or organized into Content Delivery Networks (CDNs).
These proxies are HTTP servers at their core, specialized in providing services such as caching, load balancing, and traffic filtering. A given HTTP request is processed and sometimes transformed by all the servers on the path, until it reaches the final destination, the origin server. In this paper, we refer to all of these HTTP processor hops as \textit{servers} for brevity.

An HTTP request consists of three broad sections: 1) The request line containing the request method, path, and the protocol version, 2) request headers, which are key-value pairs each terminated by the carriage return/line feed (i.e.,.~\texttt{CRLF}) byte sequence, and 3) an optional request body. 

Upon receiving a request, the server locates the requested resource based on the path provided in the request line and the \texttt{Host} header value, making correct processing of these fields crucial.

While headers can be trivially parsed thanks to the \texttt{CRLF}  deliminator, the body consists of unstructured, raw bytes. As a result, another key element of request processing is the correct parsing of the body section, which is necessary for correct interpretation of message framing (i.e., where a request starts and ends in the request buffer).

HTTP/1.1 provides two ways to encode the body and signal message framing.
\begin{enumerate}[leftmargin=*]
    \item \textbf{Content-Length encoding.} This method involves the use of a \texttt{Content-Length} header, where the value corresponds to the total body length specified in bytes. The processing of Content-Length encoded requests is straightforward; the server parses the \texttt{Content-Length} value, and consumes the specified amount of bytes from the request buffer, interpreting those bytes as the body. 
    
    \item \textbf{Chunked encoding.} The second method encodes the body in multiple \textit{chunks}. Each chunk consists of a hexadecimal value representing the length of the chunk's data section specified in bytes, followed by the said data. The end of the request is indicated by an empty chunk of length "0" that contains no data. Both the chunk length and chunk data lines are terminated by the \texttt{CRLF} sequence. This encoding scheme is particularly suitable for streaming applications, since each chunk can be individually sent, received, and processed in a streaming fashion. Also note that chunked encoding is \textit{necessary} when the full length of the body is not known at the time of processing, making it impossible to calculate the \texttt{Content-Length} value. Chunked encoding is indicated by including the \texttt{Transfer-Encoding} header with the "chunked" value in a request. 
\end{enumerate}

An important decision for proxied architectures is whether to buffer an entire request for processing its body as a whole, or whether to process and forward bytes to the next hop as soon as they are received, also known as streaming the body. This decision may be influenced by many factors revolving around the server implementation and application logic. Buffering has severe performance ramifications, roughly doubling the transfer latency for every hop that needs to wait for the entire request to arrive before sending it forward.

\subsection{HTTP Extensibility}
\label{sec:extensibility}

Beyond the standard headers defined in the specification, HTTP provides mechanisms to extend the protocol. Specifically, application owners can extend HTTP by introducing \textit{custom headers} and \textit{trailer fields.}

Custom headers follow the same structure and parsing semantics as regular headers. There are no restrictions on custom header names. Custom headers can be used for any purpose that the application owners deem appropriate.

An important requirement for the functioning of custom headers is that they must be forwarded intact by intermediate servers that do not know how to process them. This is required specifically in the RFC (see RFC 9110, Section 5.1~\cite{rfc9110}). 



Unlike what we described for custom headers, the HTTP specification does not dictate the forwarding of trailer fields by intermediate servers (Section 7.1.2 of RFC~\cite{rfc9112}). 
As a result, while some servers support trailer fields, others do not. Therefore, if trailer fields are used, users should ensure that all servers forward them correctly.

\subsection{HTTP Processing Discrepancy Attacks}
\label{sec:discrepancy_desc}

When at least two servers on the traffic path process the same HTTP request in different ways, a processing discrepancy occurs. Attackers can then abuse these discrepancies to harm web applications and their users. We collectively call such attacks \textit{discrepancy attacks}. We classify the currently documented processing discrepancy attacks into three categories based on the discrepancy vector: \textit{path confusion}, \textit{host confusion}, and \textit{request framing confusion}.

The abuse of these discrepancies then take on different names such as web cache poisoning, web cache deception, request smuggling, response queue splitting, etc., depending on the types of hazardous interactions that ensue and the resulting damage. In this work, we are not particularly concerned with these specific attack types and terminology, as they are largely influenced by the victim web application's logic. That is, the same discrepancy vector may escalate to different attacks for different vulnerable applications. Instead, we focus on the discrepancy vector and mechanism, and eliminate that mechanism with our proposed defense.

We explain each discrepancy vector in more detail below. We provide specific references for the described issues later in the related work, in Section~\ref{sec:related-work}.


\noindent \textbf{Path Confusion:} 
Path confusion attacks result from processing discrepancies in the path field of a request line.

\begin{figure}[t]
\centering
\includegraphics[width=1.0\linewidth]{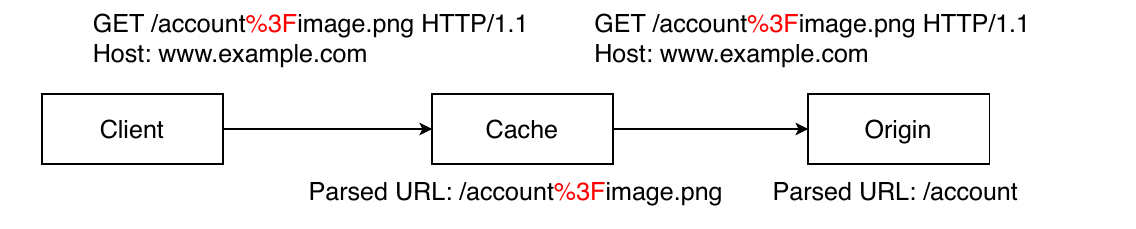}
\caption{Path confusion, leading to web cache deception.}
\label{fig:path_confusion}
\end{figure}

\begin{figure}[t]
\centering
\includegraphics[width=1.0\linewidth]{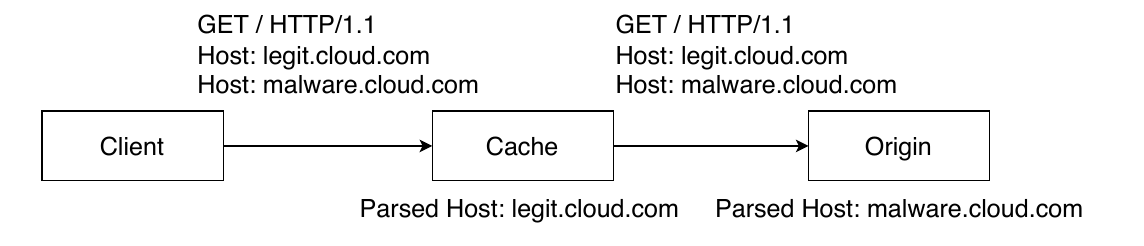}
\caption{Host confusion, leading to web cache poisoning.}
\label{fig:host_confusion}
\end{figure}

A common reason for these is the widespread use of URL representations that are designed to be accessible for Internet users, but may be difficult for proxy servers to interpret correctly. For example, consider a payment processor that exposes an "Account Details" endpoint that takes an account identifier parameter. In lieu of using a traditional filesystem path metaphor and query string representation for the resource ("account.php?id=1234"), this application may choose to employ the popular \textit{clean URL} representation ("/account/1234"). However, to an intermediate server, the interpretation of this clean URL is ambiguous; the proxy cannot know whether the client requested the "/account" endpoint with the parameter "1234", or instead an object literally named "/account/1234".

Another common issue is the existence of multiple parameter delimiter characters, each honored by a different set of servers and web application frameworks (e.g., question mark, semi-colon, sharp sign). Exacerbating the situation are tricks employed by attackers such as URL-encoding or double-URL-encoding such delimiters (e.g., "?", encoded as "\%3F", or double-encoded as "\%25\%33\%46"), exposing even more opportunities for path confusion.  

Revisiting our payment processor example, Figure~\ref{fig:path_confusion} illustrates path confusion, leading to a web cache deception attack. The attacker crafts a URL that includes an encoded question mark. The caching proxy on the way does not treat this character as a delimiter, and believes that the client is requesting an image object. The origin server, however, honors the encoded question mark as a valid delimiter, and is confused into responding with the "/account" endpoint, which contains sensitive information. The proxy stores the response in the publicly accessible cache, erroneously believing that it is a harmless image. The attacker can then trick their victims into clicking on this malicious URL via social engineering, and harvest their account information stored in the cache.


\noindent \textbf{Host Confusion:}
The destination host name, indicated by the \texttt{Host} header value, is a critical component for correct routing of requests. Host confusion refers to discrepancies in the processing of this value, which may eventually lead to serving incorrect content to the client.

Documented techniques to induce host confusion cover a myriad of tricks where the attacker includes characters that are not allowed in the standard, or encoded versions of these characters in the \texttt{Host} header. Another common cause for confusion is requests that include multiple \texttt{Host} headers with conflicting values. Similar to the path confusion scenario, servers respond to such bad \texttt{Host} headers in various ways. Some outright reject the request, others parse it in unexpected ways which lead to processing discrepancies. 

Figure~\ref{fig:host_confusion} demonstrates an attack involving multiple \texttt{Host} headers, one pointing to the intended legitimate resource on a cloud storage service, while the other is an attacker controlled storage bucket on the same cloud that serves malware. The caching proxy honors the first \texttt{Host} header it encounters, and forwards the request to the destination domain. The cloud's origin server, however, honors the second header, responding with malware. The proxy then caches the malware instead of the legitimate resource. From that point on, any client that requests "legit.cloud.com" will receive malware from the cache.


\noindent \textbf{Request Framing Confusion:}
\label{sec:sub:hrs-back} 
Request framing confusion refers to inconsistent decisions as to where a given request starts and ends. Since the parsing of the request line and header section is trivial with \texttt{CRLF} terminators, framing confusion is primarily caused by \textit{body} parsing discrepancies. 

Techniques leading to inconsistent parsing of the body include signaling Content-Length and chunked body encoding to the server \textit{simultaneously} by including both of their corresponding headers. Attackers may also mutate these headers and their values (e.g., include white-space before header names, include non-digit characters is a size field), abusing the distinct request validation (and sadly, sometimes request sanitization, a security anti-pattern) behaviors of servers. 

Listing~\ref{lst:request_smuggling_payload} depicts a payload for a popular attack, HTTP request smuggling, that results from a framing confusion. This is a maliciously crafted request that contains a \texttt{Content-Length} header and a \texttt{Transfer-Encoding} header with the value ";chunked". Notice that the \texttt{Transfer-Encoding} header's value is mangled with a semi-colon, making it invalid.

The proxy ignores \texttt{Transfer-Encoding} due to its invalid value, honors the \texttt{Content-Length} header instead, parses the body accordingly, and forwards the entire payload to the origin server as a single request. In contrast, the origin employs a more lenient parser that sanitizes the semi-colon inside the \texttt{Transfer-Encoding} header, ignores \texttt{Content-Length}, and proceeds to treat the body as chunk encoded. As a result, the origin sees two separate requests, as depicted in Listings~\ref{lst:original_request} and \ref{lst:smuggled_request}, respectively. An attacker can use such a payload to smuggle hidden requests through a proxy, bypassing web application firewall checks deployed on the proxy. In our example, the attacker leverages this to access an administrative endpoint that would normally be blocked at the proxy.

\subsection{Attacks \& Measurements in Prior Work}
\label{sec:related-work}

HTTP processing discrepancy attacks have been seen as early as 2005 with the documenting of HTTP Request Smuggling (HRS)~\cite{smuggling_old}. These attacks have re-emerged due to the increasing complexity of today's web application architectures. Researchers have recently proposed new variants of discrepancy attacks, which present assorted novel discrepancy triggers (e.g.,~\cite{httpdesync,smuggling_new, defparam, custodio1, orange_2024}).

Jabiyev et al. developed a black-box differential fuzzer, "T-Reqs," and investigated HRS within a scientific framework, discovering many more vulnerabilities impacting popular servers~\cite{jabiyev2021t}. Similarly, Shen et al.~\cite{hdiff} developed a fuzzer, named "HDiff," expanding the test surface to include host confusion and path confusion attacks. More recently, Jabiyev et al. presented another differential fuzzer, "Gudifu," focusing on gray-box guided fuzzing of servers~\cite{jabiyev2024gudifu}.

Chen et al.~\cite{host-of-troubles} documented host confusion, demonstrating how servers process the \texttt{Host} header in a non-compliant and inconsistent manner, leading to discrepancy attacks. 

Amit Klein presented an early attack that used HTTP response splitting for web cache poisoning~\cite{klein2004divide}, and more recently, James Kettle published a collection of novel techniques to poison caches~\cite{cache_poisoning, entanglement}. Nguyen et al. introduced a new class of cache poisoning attack that results in erroneous caching of error responses, making the legitimate content inaccessible, and therefore causing a denial of service~\cite{cpdos}. Liang et al. conducted a large-scale measurement of web cache poisoning vulnerabilities in the wild~\cite{liang2024internet}.

Omer Gil coined the term web cache deception (WCD) for a specialized variant of cache poisoning, where attackers exploit path confusion between a cache and the origin server to leak confidential data into a public cache~\cite{gil2017web}. Mirheidari et al. demonstrated the impact of the issue through a measurement on popular sites, also introducing new path confusion techniques~\cite{mirheidari2020cached}. Mirheidari et al. followed up this work with a large-scale Internet study, demonstrating the prevalence of the issue, as well as contributing an Internet-scale measurement methodology for WCD~\cite{mirheidari2022web}.

Although not directly related to this work, we must mention that researchers have also explored whether processing discrepancies can be repurposed in creative ways for the benefit of the Internet community. Topcuoglu et al. presented a web server infrastructure fingerprinting methodology and tool, "Untangle," which can detect individual proxy layers by relying on processing discrepancies~\cite{ndss2024untangle}. Muller et al. leveraged processing discrepancies to evade HTTP censorship~\cite{muller2024turning}.

\subsection{Defenses in Prior Work}
There exists no comprehensive solution against discrepancy attacks in academic literature, the rest of the public research domain, or the commercial security space.

AWS released "HTTP Desync Guardian," which performs various compliance validation checks on HTTP requests in order to detect common signs of request smuggling payloads~\cite{aws_guardian}. There is no evaluation of the effectiveness of this tool publicly available, and we expect it to be inherently limited due to the use of payload specific detection rules. Amit Klein described a similar "Request Smuggling Firewall" that performs various request compliance checks~\cite{smuggling_new}. This work is subject to the same limitations as above.

Cloudflare implemented a feature called "Web Cache Deception Armor" in their content delivery services, which validates origin server responses based on the \texttt{Content-Type} value, which factors into the cacheability decision. Even though this defense layer can be bypassed, it is still a practical solution against common WCD variants--albeit without a public evaluation of its effectiveness. Of course, this solution does not address any other discrepancy attack.

Beyond the above attempts at a more generic solution, efforts to defeat discrepancy attacks today focus on blocking them at a web application firewall (WAF) (e.g.,~\cite{akamai_waf}). Such WAFs typically rely on detection rules written in regular expressions that must be continuously maintained. Therefore, this approach is inherently limited by the well-established scalability and effectiveness issues plaguing signature based attack detection schemes.

Buttner et al. also proposed a WAF-like approach to detect discrepancy attacks, based on header allowlisting~\cite{buttner2021less}. This has the same limitations as WAFs that we described above. 

We must reiterate that newer HTTP protocol versions, namely HTTP/2 and HTTP/3, address certain discrepancy attacks. 
As a result, the basic form of request smuggling as we described in Section~\ref{sec:sub:hrs-back} is no longer viable in HTTP/2 or HTTP/3--but the remaining path and host confusion attack vectors remain intact. 
Furthermore, researchers have shown that the complexity involved in these protocol updates, 
introduce brand new opportunities for discrepancy attacks~\cite{kettle-h2-hrs, sec2022frameshifter}. 
In practice, HTTP/1 remains in support on virtually every server for backwards compatibility with user agents, and therefore, the protection afforded by HTTP/2 or HTTP/3 is not a barrier for attackers.

Another loosely related concept is HTTP message signatures. Cryptographically signed HTTP header sections, or the full body payload, is a common occurrence, for example, when implementing authentication schemes for REST APIs. There is also an "HTTP Message Signatures" RFC currently in the standardization track~\cite{rfc9421}. Signed messages can potentially block specific discrepancy attacks; for example, the smuggling attack previously depicted in Listing~\ref{lst:request_smuggling_payload} results in a split body payload that would fail a signature check, provided that the attacker cannot find a discrepancy to confuse the signature validation as well. That said, this protection is once again a beneficial side-effect that may block certain attacks. Message signatures neither aim to prevent discrepancy attacks, nor are they capable of providing systemic defense. Many discrepancy attacks do not require a message integrity violation. Case in point, the two attacks we described in Figure~\ref{fig:path_confusion} and Figure~\ref{fig:host_confusion} involve no changes to either the headers or body, and therefore, cryptographic integrity checks are irrelevant.

More broadly, any protection relying on point-checks for request validation is necessarily limited, since there exists no processing context during the check. In contrast, providing a comprehensive defense solution requires validation of the processing history for all server hops on the traffic path.

\begin{noerr}

\begin{listing}[t]
\caption{Framing confusion payload leading to request smuggling.
The \texttt{CRLF} sequences in the body are explicitly noted since these bytes
contribute to the Content-Length header value.}
\label{lst:request_smuggling_payload}
\begin{lstlisting}[style=httpcode]
||POST|| <</public>> ((HTTP))/1.1
[[Host:]] www.example.com
[[Content-Length:]] 51
[[Transfer-Encoding:]] ;chunked

0\r\n
\r\n
GET /admin HTTP/1.1\r\n
Host: www.example.com\r\n
\r\n
\end{lstlisting}
\end{listing}



\begin{listing}[t]
\centering
\begin{tabular}[t]{@{}p{0.48\linewidth}@{\hspace{0\linewidth}}p{0.48\linewidth}@{}}

\begin{minipage}[t]{\linewidth}
\captionof{listing}{Split request.}
\label{lst:original_request}
\hfill

\begin{lstlisting}[style=httpcode]
||POST|| <</public>> ((HTTP))/1.1
[[Host:]] www.example.com
[[Content-Length:]] 51
[[Transfer-Encoding:]]
(*@\textcolor{gray}{\ensuremath{\hookrightarrow}}@*) chunked

0
\end{lstlisting}
\end{minipage}

&

\begin{minipage}[t]{\linewidth}
\captionof{listing}{Smuggled request.}
\label{lst:smuggled_request}
\hfill
\begin{lstlisting}[style=httpcode]
||GET|| <</admin>> ((HTTP))/1.1
[[Host:]] www.example.com







\end{lstlisting}

\end{minipage}

\end{tabular}
\end{listing}

\end{noerr}

\section{Research Statement}
\label{sec:research-statement}

HTTP processing discrepancies, and therefore discrepancy attacks, have been around for a long time, taking different forms and names. As summarized above, the security community's response has largely focused on bug hunting, patching, or blocking specific attack payloads through web application firewalls and regular expression based rules. This is not a scalable approach and does not prevent damage, as the security community must reactively respond to evolving attacks after incidents.

A major contributing factor to this situation, as repeatedly emphasized by prior work (e.g.,~\cite{gil2017web, mirheidari2020cached, mirheidari2022web, jabiyev2021t}), is that discrepancy attacks result from hazardous interactions between servers that may otherwise work flawlessly in isolation. Hence, this is a problem at the intersection of security and \textit{safety} engineering. Safety issues are relatively new to the web application security community, only recently exacerbated due to the proliferation of complex proxied architectures. Finding effective solutions for safety issues is not straightforward, as there is not necessarily a faulty component to single out and fix.

Our problem statement can be summarized as such: Despite the surge in damaging discrepancy attacks, there exists no generally applicable defense. Our research statement directly follows from this problem: Can we design and implement a defense that addresses the core of this problem?

We hypothesize that communicating request processing history across all proxies present on the traffic path, in a format enabling each individual server to validate that they are indeed processing the request consistently with all previous hops, would address processing discrepancies. Discrepancy attacks are sometimes called HTTP request desynchronization in the literature, hence, it is only appropriate that we call our defense scheme \textit{HTTP Request Synchronization}.

To make our contributions practicable, we also establish a set of design goals as laid out below.

\begin{itemize}[leftmargin=*]
\item \textbf{Do not violate, or modify the HTTP protocol specification.} Any design that violates the HTTP specification exacerbates the problem by running the risk of introducing more processing discrepancies. Moreover, it is not operationally feasible to make changes to such a well-established and widely deployed protocol. Therefore, in our design, we shall only rely on the extension mechanisms provided by the HTTP standard, or otherwise augment the request without violating the specification. We shall propose a solution that only requires minor, practical changes to server implementations that can be applied by the technology vendors or consumers.

\item \textbf{Support unmodified servers.} To enable the request synchronization checks we propose, modifications to servers are inevitable. However, it may not always be possible for \textit{all servers} on the traffic path to implement our defense (e.g., there could be proxies that are not under the control of the infrastructure owner). Therefore, our solution must remain functional in the presence of oblivious servers that are unaware of the HTTP Request Synchronization scheme. This introduces two intertwined design requirements: 1) We shall not disrupt normal HTTP processing of oblivious servers, and 2) oblivious servers shall not completely undermine the security guarantees of our defense, even though some guarantees may need to be relaxed.

\item \textbf{Do not modify user agents.} We shall not require any changes to user agents, such as browsers, as this is unlikely to be attainable.

\item \textbf{Avoid performance bottlenecks.} HTTP servers employ optimizations to support the low-latency data streaming requirements of contemporary applications, such as online gaming, video conferencing, and big file uploads. This poses challenges for message security schemes, making a naive implementation that requires buffering of the entire request for cryptographic computations unacceptable--such buffering could result in over a 100\% transfer time overhead by undoing the benefits of streaming. A defense resulting in such drastic performance degradation is unlikely to be employed, no matter the security benefits. Hence, in our design, we shall not introduce performance bottlenecks that cripple streaming data flows.
\end{itemize}

The threat model for our work tracks a typical web application and Internet attacker model. The attacker's user agent has full control over the request, including any request line, header, and body. That includes the request components we introduce to enable HTTP Request Synchronization; the attacker is allowed to craft requests that contain such valid, invalid, or forged components. We assume that the servers in the traffic path are not controlled by the attacker, and finally, TLS/HTTPS is used to prevent man-in-the-middle attacks that violate traffic confidentiality and integrity.

In this work, we assume that the request body is immutable during processing by the servers. To the best of our knowledge, no typical reverse proxy use case involves modifications to the request body in flight. For example, web applications firewalls strictly operate in an allow-or-block fashion, but they do not attempt to sanitize body bytes. Similarly, any request transformation use cases we are aware of strictly operate on the URL or the header section (e.g., URL rewriting, redirection, header injection, header stripping).

Finally, we reiterate that this research is scoped for ensuring consistent request processing across a proxied architecture, which results in blocking discrepancy attacks. As we explained in Section~\ref{sec:discrepancy_desc}, we define a discrepancy attack as any harmful outcome caused by hazardous interactions between multiple servers, which may occur even when the individual servers are bug-free. However, many of the specific attacks we have described so far may also be caused by plain software bugs in individual server implementations, requiring no unsafe interaction with another server (e.g.,~\cite{apache_cve_rewrite,varnish_cve,nginx_cve}). For example, a buggy server may read the correct body payload from an incoming request, but craft an incorrect forward request with a partial body due to an implementation fault of its own, effectively resulting in request splitting. Such cases are, by definition, not discrepancy attacks; they can be detected and addressed by the server developers through established program analysis and testing processes. These are therefore outside the scope of our work. We make no claims to solve all web poisoning, request smuggling, request splitting, etc., attacks; as we have explained, our work is focused on the confusion vectors leading to discrepancies, which cannot be addressed through traditional secure software development practices. 
\section{Design}
\label{sec:design}

\begin{figure*}[!htbp]
\centering
\includegraphics[width=1.0\textwidth]{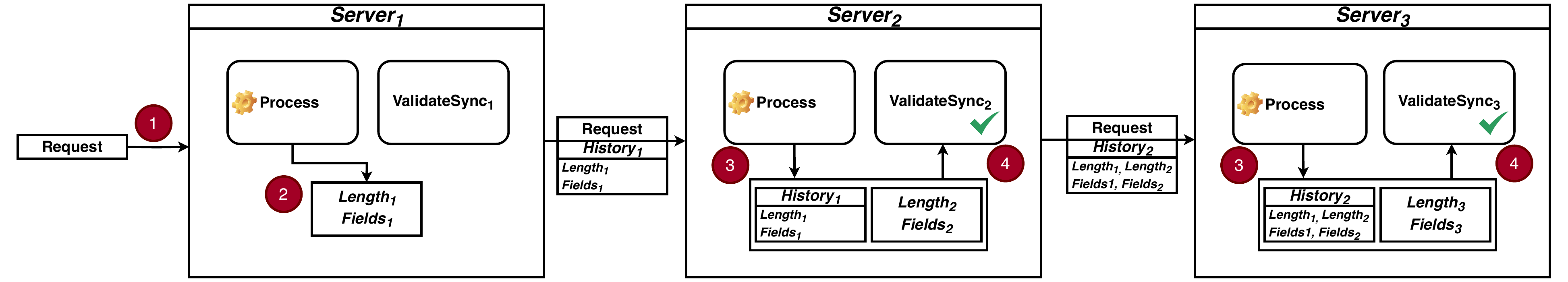}
\hfill
\caption{HTTP Request Synchronization demonstrated with three servers.}
\label{fig:methodology_example}
\end{figure*}

Attacks may abuse processing discrepancies resulting in path confusion, host confusion, or request framing confusion, as discussed above. Therefore, our defense scheme must ensure consistent processing (i.e., \textit{synchronization}) of request components that factor into path, host, and framing decisions. In this section, we initially provide an abstract description of HTTP Request Synchronization that applies to any request component, and any body encoding scheme. Implementing this scheme in a real HTTP server introduces additional engineering challenges; we cover these in detail later in Section~\ref{sec:implementation}.

We can intuitively describe HTTP Request Synchronization as follows: Servers augment each request with a description of how they processed it before sending the request forward. Consequently, requests accumulate a history of processing information for each hop they traverse. Servers can then consult this history, compare their own processing behavior to all previous hops, and apply a suitable security policy.

We define these interactions pseudo-formally below.
\begin{itemize}[leftmargin=0pt]
    \item[] \textit{$\bm{Server_i}$} is a server on the traffic path, $1 \leq i \leq n$, where $Server_1$ is the first server receiving the request from the user agent, and $Server_n$ is the origin server.
    \item[] \textit{$\bm{Length_i}$} is the request length honored by ${Server_i}$.
    \item[] \textit{$\bm{Fields_i}$} is the collection of request components (e.g., the path, header field-value pairs) that require synchronized processing, honored by ${Server_i}$.
    \vspace{0.3em}
    \item[] \textit{$\bm{History_i}$} = \small $\twopartdef {\{Length_j,Fields_j\}\,|\,\forall j \in \{1, \dots, i\}} {i > 0} {\emptyset} {i = 0}$
    \normalsize
    \item[] \textit{$\bm{ValidateSync_i}$} is a function defined for ${Server_i}$, which validates request synchronization according to the server's \textit{security policy}, where $f:(Length_i, Fields_i, History_{i-1}) \to \{Valid, Invalid\}$.
\end{itemize}

Note that our use of the term "honored" in the above description is not arbitrary. We define the honored value as the ultimate value that the server uses in its processing of the request. Contemporary servers have complex request processing pipelines, and the honored value of a request component may be different from the initially parsed value--this is in fact a contributing factor to processing discrepancies. How and where our defense intercepts this ultimately honored value is a server-dependent implementation detail.

Consequently, $Server_i$ performs the following actions.
\begin{enumerate}
    \item Receives a request augmented with $History_{i-1}$.
    \item Processes the request to assemble $Length_i$ and $Fields_i$.
    \item Executes $ValidateSync_i(Length_i, Fields_i, History_{i-1})$ to validate the processing according to its security policy.
    \item If the outcome is valid, assembles $History_i$, and augments the forward request with it. If the outcome is invalid, terminates the connection.
\end{enumerate}

In the base, and most common use case, $ValidateSync$ can be realized with two simple checks:
\begin{enumerate}
    \item $Length_i$ for all $i$ must be equal.
    \item $Fields_i$ for all $i$ must be equal, further defined as strict bitwise equality for all field-value pairs.
\end{enumerate}
However, in practice, some web applications may require intentional modifications to sensitive fields during proxy processing. For example, a load balancer may rewrite the destination host. This is why we must let $ValidateSync_i$ be a custom validation routine, defined for each $Server_i$ based on its own security policy. In this way, servers can implement complex validation checks over the entire history, confirming that field modifications happen as expected, on the expected hop.

Figure~\ref{fig:methodology_example} depicts an example with three servers.
\begin{enumerate}
    \item ${Server_1}$ receives a request from the user agent.
    \item ${Server_1}$ is the first processor on the path; therefore, there is no history available, and no synchronization validation necessary yet. The server processes the request, initializes \textit{${History_{1}}$} with \textit{${Length_{1}}$} and \textit{${Fields_{1}}$}, crafts a forward request that includes \textit{${History_{1}}$}, and sends the request to ${Server_2}$.
    \item ${Server_2}$ processes the request and validates the received \textit{${History_{1}}$}, including \textit{${Length_{1}}$} and \textit{${Fields_{1}}$}, by comparing against \textit{${Length_{2}}$} and \textit{${Fields_{2}}$}.
    \item Assuming that the validation succeeds, ${Server_2}$ assembles \textit{${History_{2}}$} by supplementing \textit{${History_{1}}$} with \textit{${Length_{2}}$} and \textit{${Fields_{2}}$}, builds the forward request that includes \textit{${History_{2}}$}, and forwards the request to ${Server_3}$.
    \item ${Server_3}$ performs the validation by comparing the values in \textit{${History_{2}}$} with \textit{${Length_{3}}$} and \textit{${Fields_{3}}$}. Since this is the origin server, there is no further forward traffic. If validation succeeds, ${Server_3}$ returns the appropriate response to be delivered back to the user agent.
\end{enumerate}

This design guarantees the synchronized processing of requests on all hops, assuming that all servers support our scheme. In cases where there are oblivious servers present on the path, provided that they do not destroy the history embedded in the request, we can still provide security guarantees, albeit looser ones. Specifically,
over contiguous hops of servers equipped with our scheme, request synchronization is guaranteed. If multiple such server ranges are separated by oblivious hops, our scheme may still be able to detect a processing discrepancy impacting one of the oblivious hops if followed by a server using the scheme. This is enabled because of the forwarded processing history (e.g., a request body split in an oblivious server would result in a length mismatch that can still be detected later in the processing). Our security guarantees end at the final hop that supports our scheme, all further processing is exposed to attacks.
\section{Implementation}
\label{sec:implementation}

We now present a concrete implementation based on our design. On each server, we need to construct a $History$, which contains $Fields$ and $Length$, and validate the request synchronization. All documented discrepancy attacks in literature involve path confusion, host confusion, and request framing confusion vectors all caused by body length confusion. Thus, our implementation covers three specific request components: The path in the request line, the \texttt{Host} header value (i.e., mapping to $Fields$), and the body length (i.e., mapping to $Length$). If circumstances change in the future, this can be trivially extended to cover more components.

The engineering challenges for constructing $Fields$ and $Length$ differ. $Fields$ can be constructed in a relatively straightforward manner by analyzing the request line and headers at the time the request is forwarded or terminally processed at the origin. In contrast, \textit{Length} can be encoded in two different ways depending on whether Content-Length encoding or chunked encoding is used. More importantly, the total body length is not always known at the time a forward request is sent--if the server performs request streaming, the request line and headers will be sent before the complete body is received from the previous hop.

Because of this distinction, we tackle the implementation of $Fields$ and $Length$ in their respective sections below. Throughout these sections, we revisit the design targets defined in Section~\ref{sec:research-statement} to ensure practicality of our implementation. Finally, in Section~\ref{sec:complete-implementation}, we fill in the missing pieces and present the complete implementation of HTTP Request Synchronization.

\subsection{Path and Host Name}

The implementation for the path and \texttt{Host} header is straightforward. At the time of forwarding the request, we have full visibility into these values. We only need to perform the synchronization check and propagate these fields to the next hop.

We propagate this information by embedding it into the forward request using a custom header, named \defenseheader, which contains the $Fields$ values that constitute part of $History$. We define \defenseheader as a JSON object, where each key represents a field, and its corresponding value is a list containing the field's values over all server hops.

This implementation aligns with our design goals. Using a custom header ensures compliance with the HTTP protocol specification. It also guarantees that the propagated information will not be stripped away by oblivious servers since the specification mandates forwarding of unrecognized custom headers. Note that streaming concerns are not relevant here, as the request line and header section are processed identically regardless of whether the server performs streaming.

Upon receiving the request, the server processes it normally, until our implementation intercepts it, allowing us to access the path and host name. We first check whether \defenseheader is included among the headers. If it is not, this server must be the first processor and there is no $History$ available. We initialize \defenseheader using the current field values and perform no validation.

Otherwise, if \defenseheader is present in the received request, we validate the processing of the current path and host against the received $History$, via the $ValidateSync$ function. After successful validation, we update \defenseheader with the current path and host name to update the $History$, and give control back to the server to forward the request.

As mentioned before, the default and most commonly applicable validation strategy for these fields is a strict bitwise string equality check. However, some applications such as load balancing may involve legitimate modifications to either the path or the host name. Therefore, our implementation allows operators to define a custom validation routine over the provided list of field values in $History$, allowing them to check for intended value transitions and modifications.

In fact, for the final checks performed on origin servers, it may be more appropriate to implement the more complex validation routines at the web application layer, instead of the HTTP server layer. Since the application owners have full visibility into the security policy requirements, they can implement comprehensive synchronization validation as part of their application's regular input validation gate. There is no special implementation effort needed on our part to support such checks; web application frameworks already expose headers, and therefore \defenseheader, to the application logic, allowing application owners to trivially implement more validation over it. We demonstrate such a scenario that involves origin application layer validation of $History$ in one of our case studies later in Section~\ref{sec:casestudy}.

\subsection{Body Length}

\begin{figure}[!t]
\centering
\includegraphics[width=0.79\linewidth]{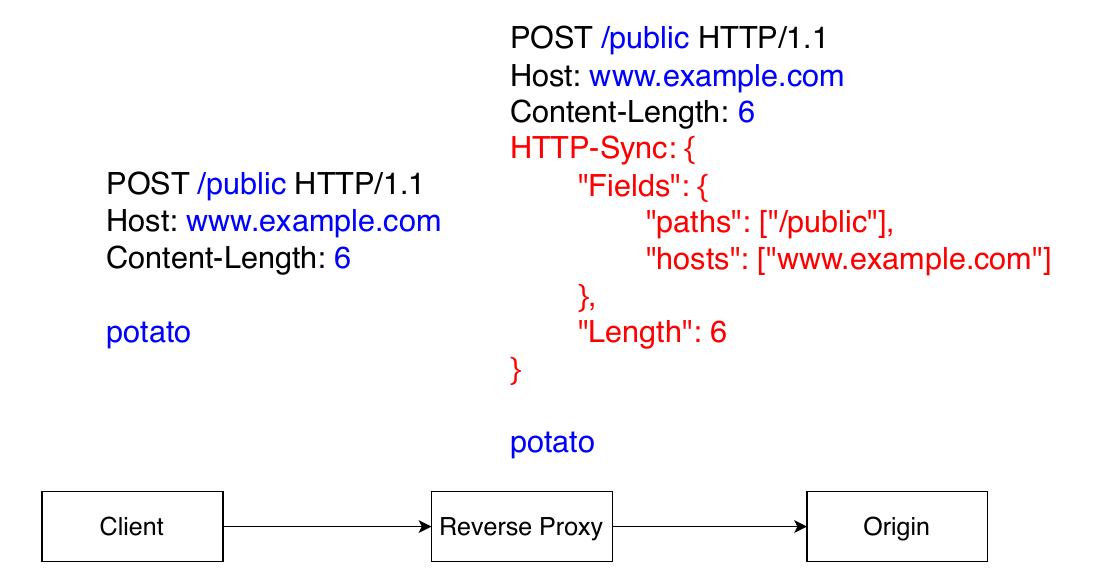}
\caption{Implementation with Content-Length encoded body. Blue text indicates the honored values. Red text indicates our enhancements to the original request.}
\label{fig:request_smuggling_defense}
\end{figure}

The implementation for body length is more complex, primarily because servers may not know the exact body length at the time of forwarding. We first address the straightforward case, where the request body is Content-Length encoded.

Recall that for Content-Length encoded requests, the body length is signaled with the \texttt{Content-Length} header. The server processing a request immediately knows the body length as it receives the headers from the previous hop, without waiting to receive any body bytes. That is, regardless of whether the server performs buffering or streaming, request framing information is available for our implementation to intercept before a forward request is initiated, just like in our processing of path and host name above. 

Therefore, using a custom header to propagate the body length is also viable in this case. The honored \texttt{Content-Length} value directly maps to \textit{Length}, the remaining part of the \textit{History}, and we include this value in \defenseheader like before.

While the previously described processing principles for \defenseheader remain the same, we have the opportunity to apply an optimization thanks to the fact that $Length$ must be identical in all hops, unlike $Fields$ that may change. Specifically, it suffices to perform the synchronization on a hop-by-hop basis, where every server validates that $Length_i$ is an identical integer to the previously processed value $Length_{i-1}$. That is, there is no need to keep a complete record of all $Length$ history, and we only include the last honored $Length$ in \defenseheader. If future use cases or attacks necessitate different validation requirements, this optimization can be trivially undone.

To summarize everything so far, Figure~\ref{fig:request_smuggling_defense} depicts how a generic reverse proxy implementing HTTP Request Synchronization processes a request with a Content-Length encoded body. The proxy receives a basic request from the user agent, processes it as usual, initiates \defenseheader, and includes it in the forwarded request to the origin. The origin server then validates the path, host name, and length values it honors against the values it extracts from \defenseheader.

Requests using chunked encoding, however, pose a challenge. Recall that in chunked encoding, body length information is encoded into each individual chunk, and the total body length can only be computed once all chunks are received at the server. However, servers that implement request streaming can start data transmission as soon as they receive a header section, without waiting for the body, and thus without knowledge of the total body length. Using \defenseheader for propagating \textit{Length} fails in this scenario, because by the time our header is sent, the body length is still unknown.

A naive workaround would be to force buffering of the request body before initiating the forward transfer. This violates our design goal of avoiding impractical performance bottlenecks. Hence, we need an alternative approach. Below, we present two implementation options to tackle this challenge.

\textbf{Option \#1: Length in Trailer Fields.}
Recall from Section~\ref{sec:extensibility} 
that RFC 9110 defines trailer fields that can be sent after a chunked body, in order to support forwarding additional data that only becomes available after the processing of the body. This is a perfect fit to address the problem at hand via a standard HTTP extension mechanism. 


Initially, trailer fields appear to be the ideal solution to our problem. Unfortunately, as we discussed in \ref{sec:extensibility}, the RFC does not mandate the preservation of trailer fields when a server removes chunked encoding. Whether to retain the trailer is left as a design choice for the server. To check how this mechanism is implemented in practice, we tested five popular server technologies (Apache, NGINX, HAProxy, Varnish, and Cloudflare) to determine how they work with trailer fields. We observe that only one of the servers, HAProxy, retains trailer fields, while the rest drop them after processing the request.

This is a severe limitation. We conclude that while trailer fields provide a clean and effective implementation option, they are only viable when the infrastructure owner has full control over the traffic path, and can therefore specifically deploy servers that preserve trailers. This may still be a viable implementation in some scenarios, but it is not a general solution otherwise. Furthermore, this violates our design goal of supporting oblivious servers. All in all, we seek an alternative implementation option that fares better.

\textbf{Option \#2: Length in Chunk.}
Our second implementation option injects the total body length into the chunked body stream, in its own dedicated chunk. Specifically, we create a regular chunk that carries as its payload a custom \texttt{Length: <body-length>} field, and insert this immediately before the terminating chunk.

From a server's perspective, this custom field appears to be part of the regular body, and it is processed as such. In other words, all servers, oblivious or not, are guaranteed to preserve the information we add, since they are expected not to mangle or truncate a request's body.

Another way to think about this implementation is that we emulate the properties of a trailer field (i.e., custom metadata delivered at the end of a request), by extending the request payload. The implementation manages the injection and removal of this payload as the request is intercepted for validation, ensuring that it is not erroneously passed to the web application layer on the origin server. This meets all of our goals. We are fully compliant with the HTTP protocol, merely providing a custom extension mechanism that overlays the raw body bytes, without interfering with the normal HTTP processing. This mechanism is guaranteed to survive oblivious servers, since the body is immutable. Finally, our extension has no negative effect on streaming.

\begin{figure}[t]
\centering
\includegraphics[width=0.79\linewidth]{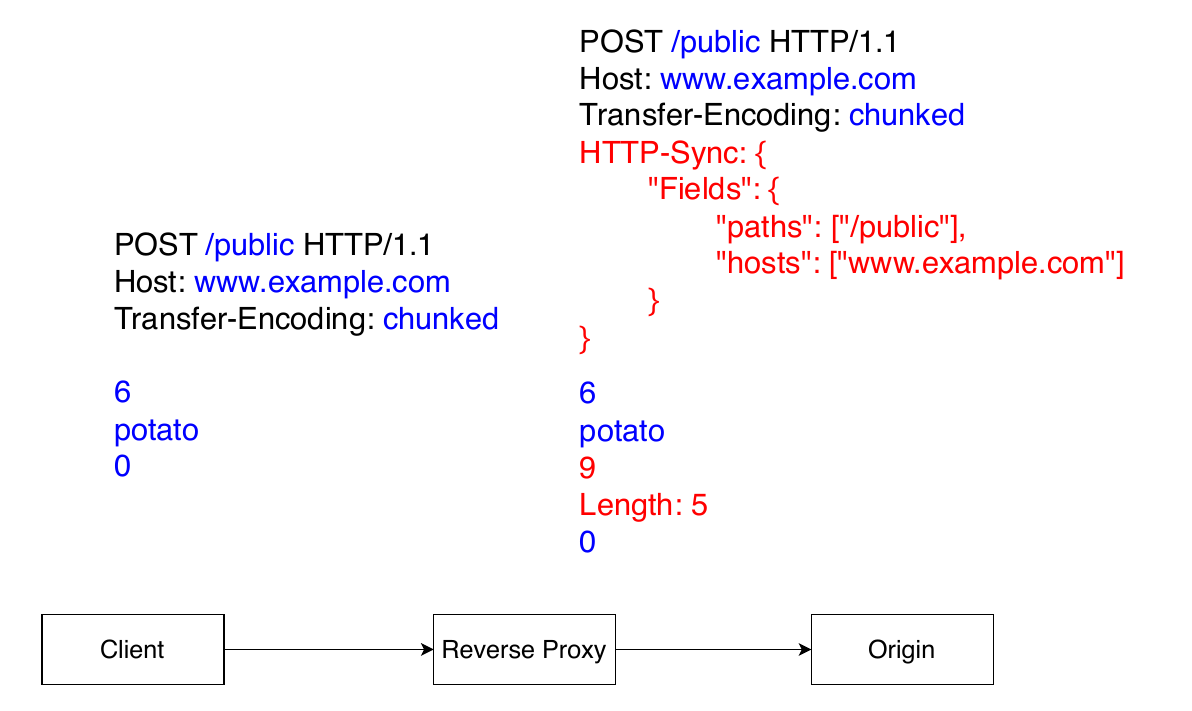}
\caption{Implementation option \#2 for requests with chunked body encoding. Total body length is injected into the stream in the penultimate chunk.}
\label{fig:embedding_to_body}
\end{figure}

Figure \ref{fig:embedding_to_body} illustrates this implementation with a streamed chunked body. We use \defenseheader to convey the host and path as before, and we inject the body length into the stream in the penultimate chunk. The processing and validation of the request is identical to the previous cases discussed.

In this work, all evaluation and source code artifacts use implementation option \#2 due to its general applicability.

\subsection{History Authentication and Integrity}
\label{sec:complete-implementation}
Our implementation is missing authentication and integrity checks on our added $History$.
The custom fields we introduce are not immune to discrepancy attacks themselves. For example, if an attacker identifies an HTTP parser discrepancy that allows them to craft malicious requests that result in incorrect processing of $History$, all guarantees are lost. Similarly, we do not make the assumption that the first processor on the traffic path is aware that it is indeed the first hop, and therefore an attacker may attempt to craft a malicious request that already includes a forged $History$ before any processing, invalidating the basis for our security claims. 

To defend against such attacks, or non-malicious, but unsafe interactions, we employ a hash-based message authentication code (HMAC). The HMAC is computed over the string representation of the value of \defenseheader, providing the necessary cryptographic authentication and integrity protection. The HMAC value is included in another custom header, \defenseheaderhmac, verified at every hop, and subsequently updated with all changes to $History$.

An inevitable operational overhead here is key sharing between the servers; a keyless cryptographic digest is not appropriate for the task, as authentication is essential. We do not aim to provide a key management solution in this research. We recognize that this is a real limitation of our approach, albeit a recurring one in proxied web architecture deployments. These deployment though, are likely to have a solution already employed by the infrastructure owners for their other secret management and distribution needs, such as a cloud key vault. Note that there is no HMAC operation performed by the client originating the request; legitimate user-agents do not need access to the key.

\subsection{Summary}
We summarize how HTTP Request Synchronization interacts with a simple HRS attempt. We later provide more complex scenarios taken from real-life attacks in Section~\ref{sec:casestudy}. 

Recall the HRS attack defined in Listing~\ref{lst:request_smuggling_payload}, where the reverse proxy treats a request as using Content-Length encoding, but the origin processes the same request with chunk encoding, resulting in the smuggling of a request that can access the forbidden administrator endpoint on the origin. 

\begin{figure}[t]
\centering
\includegraphics[width=0.79\linewidth]{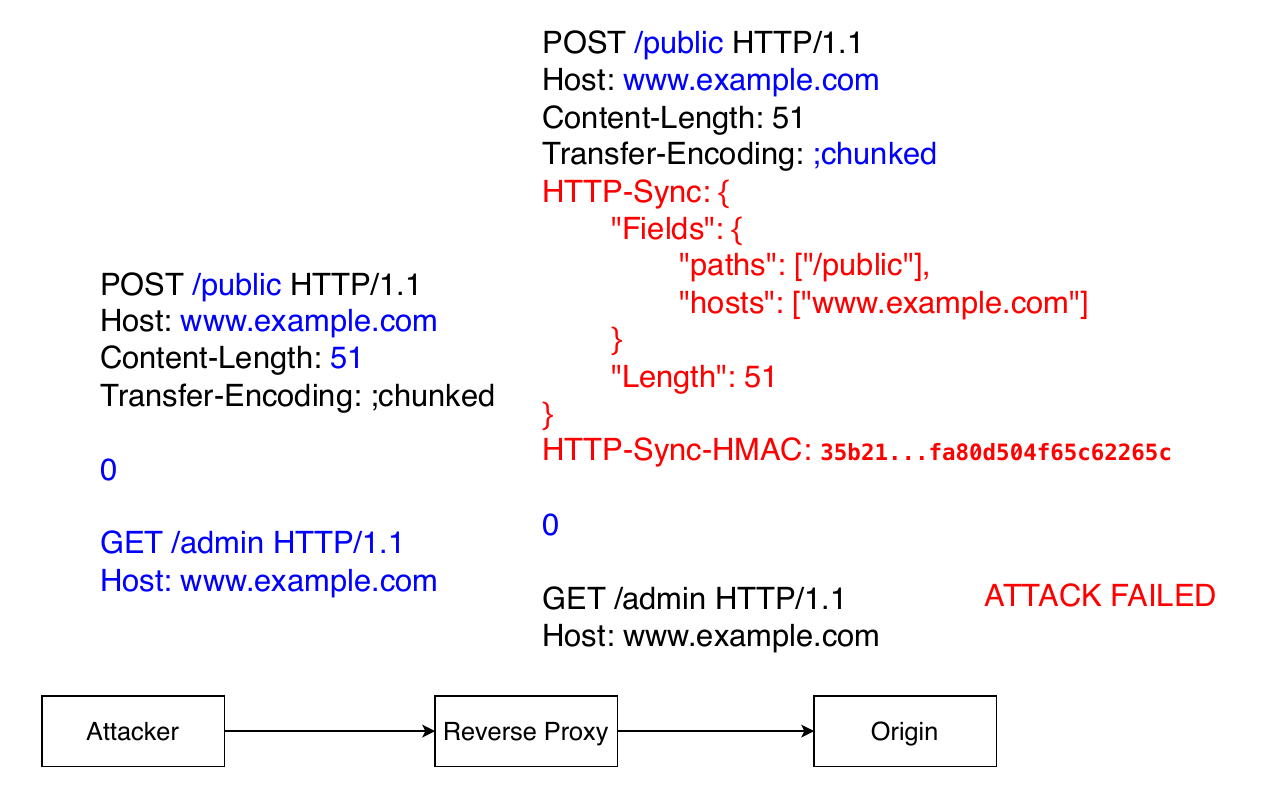}
\caption{HTTP Request Synchronization prevents HRS.}
\label{fig:smuggling_defense}
\end{figure}

Figure~\ref{fig:smuggling_defense} shows how our implementation blocks this attack. The reverse proxy inserts the processing information into \defenseheader, indicating that the body length is 51. At the origin server, the body length is instead interpreted as 0. When the origin server attempts to validate request synchronization, it compares 51 and 0, detecting a discrepancy. This results in the immediate termination of the connection, preventing HRS.

\section{Evaluation}
\label{sec:evaluation}

We first anecdotally describe our implementation process and the level of effort required for server modifications, including the challenges we encountered during the process. Then, we measure the performance impact for each server. Table~\ref{tab:servers} in the Appendix lists these servers and their respective versions.

\setlength{\tabcolsep}{3pt}

\begin{table*}[!ht]
\scriptsize
\centering
\caption{Benchmark results for original servers (i.e., $Original$, or $O$), servers that are modified by our design (i.e., $Modified$, or $M$), and a Cloudflare worker that forwards requests as-is ($Baseline$, or $B$). Mean RTT and percentage change are reported. All values in milliseconds (ms). 
\label{tab:benchmark}}
    \begin{tabular}{lclccccccccccr}
\toprule
 \multicolumn{1}{c}{} & & & \multicolumn{2}{c}{Apache httpd} & \multicolumn{2}{c}{HAProxy} & \multicolumn{2}{c}{NGINX} & \multicolumn{2}{c}{Varnish} & \multicolumn{1}{c}{Cloudflare} 
\\ 
\midrule
 \multicolumn{1}{c}{Framing} & \multicolumn{1}{c}{Chunk} &  \multicolumn{1}{c}{Body}  & \multicolumn{1}{c}{O/M} & \multicolumn{1}{c}{\%} & \multicolumn{1}{c}{O/M} & \multicolumn{1}{c}{\%} & \multicolumn{1}{c}{O/M} & \multicolumn{1}{c}{\%} & \multicolumn{1}{c}{O/M} & \multicolumn{1}{c}{\%} & \multicolumn{1}{c}{O/B/M} 
 \\ 
\midrule

        NA & NA & 0 Kb & 2.43 / 2.72 & 11.93 & 2.17 / 2.34 & 7.83 & 2.17 / 2.40 & 10.60 & 2.34 / 2.50 & 6.84 & 54.96 / 74.18 / 44.92 & 
        \\
        \midrule
        
        CL & NA & 100 b & 2.46 / 2.68 & 8.94 & 2.15 / 2.32 &  7.91 & 2.13 / 2.36 & 10.80 & 2.28 / 2.47 & 8.33 & 50.79 / 71.26 / 42.68 & 
        \\
        
        CL & NA & 100 Kb & 2.64 / 2.89 & 9.47 & 2.32 / 2.52 &8.62& 2.31 / 2.54 & 9.96 & 2.46 / 2.61 & 6.10 &  113.28 / 148.85 / 86.01 & 
        \\
        
        CL & NA & 1 Mb &  3.80 / 4.06 &6.84& 3.76 / 3.84  & 2.13 & 3.86 / 3.99 & 3.37 & 3.72 / 3.82 & 2.69 & 213.45 / 353.57 / 323.87 & 
        \\
        
        CL & NA & 10 Mb &  19.71 / 20.76 &5.33& 14.51 / 15.23 & 4.96 & 13.60 / 15.41 &13.31& 13.09 / 14.50 &10.77& 633.46 /   1157.39 / 1362.11 & 
        \\
        
        \midrule
        
        TE & 100 & 100 b &  2.45 / 2.66 & 8.57 & 2.12 / 2.38 & 12.26 & 2.10 / 2.34 & 11.43 & 2.27 / 2.48 &9.25& 53.79 / 83.67 / 44.20 & 
        \\
        
        \midrule
        
        TE & 100 & 100 Kb & 10.20 / 10.37 & 1.67 & 13.26 / 13.49  &1.73& 10.78  / 11.58 &7.42& 9.88 / 10.04 &1.62& 123.69 / 169.59 / 102.05 & 
        \\
            
        TE & 1000 & 100 Kb & 3.48 / 3.75 &7.76& 2.86 / 3.09  &8.04& 2.99  / 3.20 &7.02& 3.40 / 3.70 &8.82& 119.13 / 139.18 / 90.54 & 
        \\

        TE & 10000 & 100 Kb & 2.92 / 3.17 & 8.56 & 2.56 / 2.77 & 8.20 & 2.62 / 2.86 & 9.16 & 2.73 / 2.95 &8.06& 114.26 / 146.01 / 85.63 & 
        \\
        
        \midrule

        TE & 100 & 1 Mb & 49.74 / 50.33 &1.19& 49.59 / 56.74 &14.42 & 49.5 / 50.15 & 1.31 & 88.46 / 91.03 & 2.91 & 242.19 / 361.62 / 346.12 & 
        \\

        TE & 1000 &1 Mb & 13.94 / 14.16 & 1.58 & 15.86 / 16.07 & 1.32 & 15.9 / 16.4 & 3.14 & 13.75 / 15.14 &10.11& 218.83 / 380.82 / 329.42 & 
        \\

        TE & 10000 & 1 Mb & 7.15 / 7.70 &7.69& 7.28 / 8.25 &13.32& 7.62 / 8.03&5.38& 7.00 / 7.70 & 10.00 & 215.25 / 375.02 / 329.25 & 
        \\

        \midrule

        TE & 100 & 10 Mb & 348.04 / 352.00 & 1.13 & 404.89 / 416.54 & 2.88 & 367.5 / 375.4 & 2.15 & 874.76 / 902.55 & 3.17 & 1566.67 / 1704.50 / 2465.77 & 
        \\

        TE & 1000 & 10 Mb & 67.72 / 69.00 & 1.89 & 69.17 / 75.90 &9.73& 67.94 / 70.24 & 3.38 & 130.36 / 132.38 &1.55& 569.22 / 1140.72 / 1431.33 & 
        \\

        TE & 10000 &10 Mb & 38.92 / 40.05 & 2.90 & 39.71 / 39.79 & 0.20 & 39.4 / 40.5 & 2.79 & 41.95 / 43.07 &2.67 & 675.69 / 1346.53 / 1400.50 & 
        \\
        \midrule
        \bottomrule

    \end{tabular}
\end{table*}

\normalsize

\subsection{Server Modifications}

For each server, except Cloudflare, we systematically followed three phases:
\begin{enumerate}[leftmargin=0px]
    \item[] \textbf{Discovery.} We manually explored the documentation and code repositories to identify target modules implementing the request processing functionality we are interesting in intercepting.
    \item[] \textbf{Implementation.} We revisited the potential targets identified during the discovery phase to verify whether they are viable points for implementing our design. This phase involved numerous trial-and-error attempts and human-driven dynamic testing before finalizing the implementation. 
    \item[] \textbf{Testing.} We manually created test payloads including both benign, standards compliant requests, and attack payloads, with different sizes and framing types. We manually tested the servers to both verify that the defense works as intended, and to ensure that we do not cause regressions in normal server functionality.
\end{enumerate}

Being that Cloudflare is a CDN technology, we had no source code access which necessitated a different implementation strategy. We instead used Cloudflare Workers, Cloudflare's edge compute technology that allows users to deploy serverless code directly on edge servers for custom request processing capabilities. The discovery phase consisted of learning the Cloudflare Workers APIs.

Implementation and functional testing were both manual efforts by a single PhD student. We did not perform any static or dynamic automated program analysis in this work to enable automated instrumentation. We consider that direction outside the scope of this work. Our modifications are not complex, and can be applied by the server vendor's expert developers in a straightforward manner. 

For each server, we spent approximately three weeks in the discovery phase, one week in the implementation phase, and two days in the testing phase. Since the process involved trial-and-error attempts, these are only approximations.

\textit{Discovery} was the most challenging phase due to the size and complexity of server projects. Each server has its own implementation, coding style, repository organization, and architecture representation we must understand in order to identify the correct locations to intercept the "honored" values in complex processing pipelines. Each server we worked on had its own peculiarities. For example, Apache httpd uses a bucket brigade system to manage streams, where data is broken into smaller units called buckets, while NGINX employs buffers to temporarily store data in memory and chains, which are linked structures that allow efficient handling and streaming of data between buffers. In most of these cases, server documentation was limited, which added to our challenges.

The \textit{implementation} phase was relatively straightforward compared to discovery, since we could reuse the code snippets we had implemented for other servers. However, implementation was not without its own challenges. For example, our HAProxy implementation passed the initial functional tests; however, when we increased the body length during further testing, we found that request payloads that exceed a certain length trigger a different code path. We encountered many such minor issues that we detected by systematic testing which required us to adapt our implementation or relocate our request interception logic.


Once again, we emphasize that implementing robust, optimized, production quality code that adheres to the original servers' architecture and engineering style was not a goal of our work. We present this anecdotal description only to demonstrate that fully functional proof-of-concept implementations can be achieved with reasonable manual effort by an inexperienced implementer. We expect that an in-house developer of a server vendor that is interested in implementing our scheme would find it a quick, low-cost effort.   

\subsection{Performance Impact}


We measured the performance impact of our scheme with all five implementations by comparing the traffic round-trip time (RTT)--the time it takes for a request to travel from the client to the server and a response to come back--with that of unmodified servers. The modified servers are the full implementations of our defense methodology and are capable of preventing all types of attacks.

Our test infrastructure consists of an HTTP benchmarking tool\footnote{https://github.com/six-ddc/plow} acting as the client, our modified HTTP servers as reverse proxies, and a WSGI/Flask deployment as our origin server. Since WSGI/Flask is the last server in the communication path, we only implemented the verification part of our design, meaning it can check and validate incoming requests according to the design we propose. The benchmark tool we used required modification because its original version did not have support for varying chunk length. We deployed this test setup on our local infrastructure, except for the case involving Cloudflare, where only the client and the origin were deployed on the local infrastructure.

In all experiments, we used a POST request with a body. We tested the impact of both body encoding types on the performance. For Content-Length encoding tests, we varied the request body size between 0 and 10MB. For chunked encoding tests, we used the same overall body size, but also tested each case with chunk counts between 100 and 10000. Overall, we ran 15 experiments, where each experiment was run 1000 times; we report the average of these runs for each experiment in Table~\ref{tab:benchmark}. Note that the time does not increase with the number of chunks, because the additional work that we introduce is just one summation operation per chunk.

We conclude that these results are very promising, demonstrating that the solution is feasible. At the same, we acknowledge that throughput requirements vary by application, and very busy production platforms with low-latency requirements may see a more noticeable impact in their particular setup, especially when many layers of proxies are layered back to back. It is not feasible to draw conclusions on this front without access to a real-life application with real traffic patterns. Notwithstanding the data we present, we expect an expert implementation taking into account the architecture and implementation quirks of these servers may further improve the performance. The seemingly large fluctuations in percentage values are a product of the scale of time being measured. All absolute RTT values are measured in milliseconds, resulting in small fluctuations being overrepresented as a percentage.

The Cloudflare implementation was the outlier, with a more severe performance impact. Upon further investigation, we discovered that the overhead is largely caused due to the introduction of the Cloudflare Workers layer, even without our validation operations. Note that, as depicted, fluctuations between experiment runs were higher, as performance depends on many different factors on the Internet. This is not an ideal outcome for our work, but given that Workers is a technology used in real-life applications, application owners that already incur the overhead for other uses may also implement our scheme at a lower cost. For other uses, it appears that integrating HTTP Request Synchronization in a lower layer of processing by the CDN vendor would be more appropriate.
\subsection{Security}
\label{sec:casestudy}

\begin{figure}[t]
\centering
\includegraphics[width=0.79\linewidth]{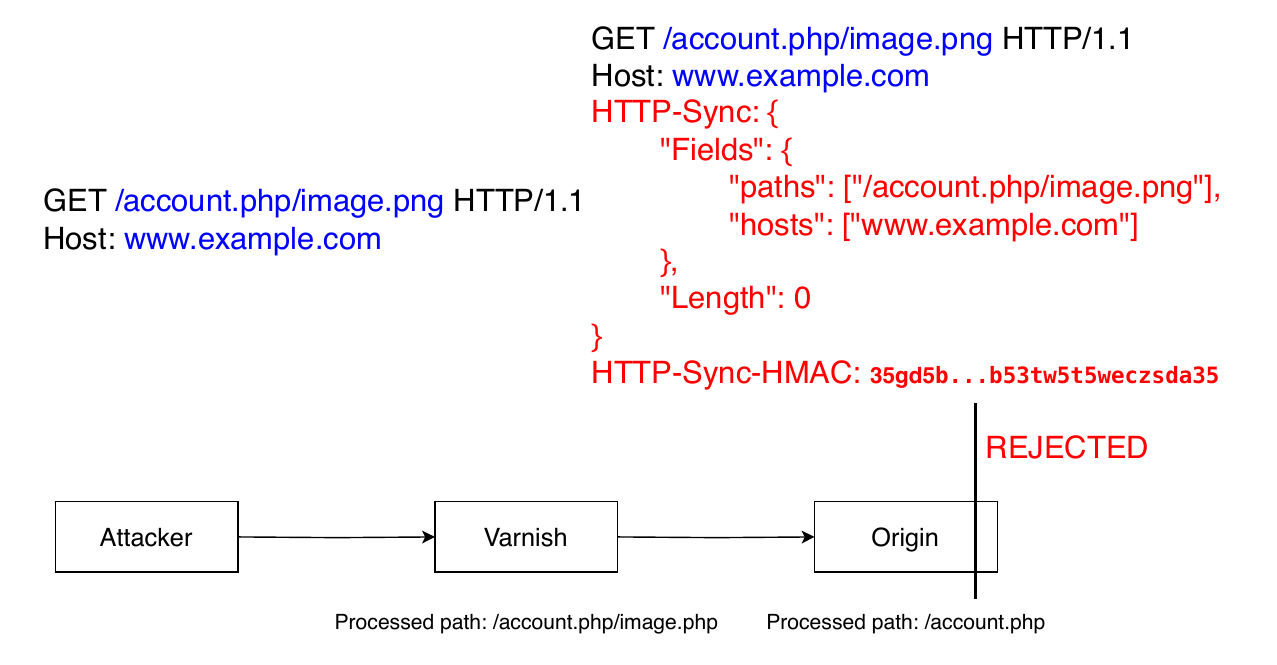}
\caption{Web Cache Deception Path Confusion Attack prevented.}
\label{fig:case_study_1}
\end{figure}

The defense we described in this work is secure by definition, without relying on probabilistic outcomes. That being said, demonstrating these security capabilities systematically is challenging. First, there is no ground truth for discrepancy attacks, but more importantly, harvesting vulnerability records such as CVEs associated with discrepancy issues and replaying them is not viable. CVEs describe a particular parsing or processing problem with a given server technology, while a full attack requires pairing at least two server technologies that have vulnerabilities that fit into the end-to-end attack scenario in just the right ways. This crucial contextual information is not captured in CVEs, let alone functional proof-of-concept exploits. 

Therefore, we instead manually analyze a selection of 34 CVEs from 2024 and 2025 that contribute to discrepancy attacks, and recreate detailed exploitable scenarios inspired by them. The CVEs we work with represent all discrepancy attack types scoped for our work. We list these CVEs in the Appendix in Table~\ref{tab:cves}.

Below, the describe this exercise in 3 distinct case studies, and empirically demonstrate that HTTP Request Synchronization defeats every attack. In all case studies, we have one or two reverse proxies, and the WSGI/Flask web application framework for the origin. We reproduce the relevant vulnerabilities in source code and implement HTTP Request Synchronization into all referenced technologies. 

\subsubsection{Case Study 1: Path Confusion}

Our first case study is drawn from Omer Gil's web cache deception~\cite{gil2017web}. In this attack, the cache server and origin disagree on the request path. Specifically, the cache server, Varnish in our example, parses the request in Figure~\ref{fig:case_study_1} and identifies the resource as "/account.php/image.png", a static PNG file. The web application framework parses the same request and serves "/account.php" instead, treating the rest as a path parameter. Varnish then caches the personal information served to the victim from "/account.php" under the path that appears to be an image.

Figure~\ref{fig:case_study_1} illustrates how we prevent this attack. Once Varnish receives the request, our design adds the full path field, including ".../image.png" into \defenseheader. The request prompts the origin to process "/account.php", but our defense compares the path received with "/account.php", observes the mismatch, and rejects the request.

\subsubsection{Case Study 2: Host Confusion}

Chen et al. discovered a host confusion between Varnish and NGINX~\cite{host-of-troubles}. They found that Varnish ignores a malformed absolute URI in the request line, and reads the host from the \texttt{Host} header. NGINX, instead, leniently parses the malformed URI and takes the host from there, ignoring the \texttt{Host} header.

\begin{figure}[t]
\centering
\includegraphics[width=0.79\linewidth]{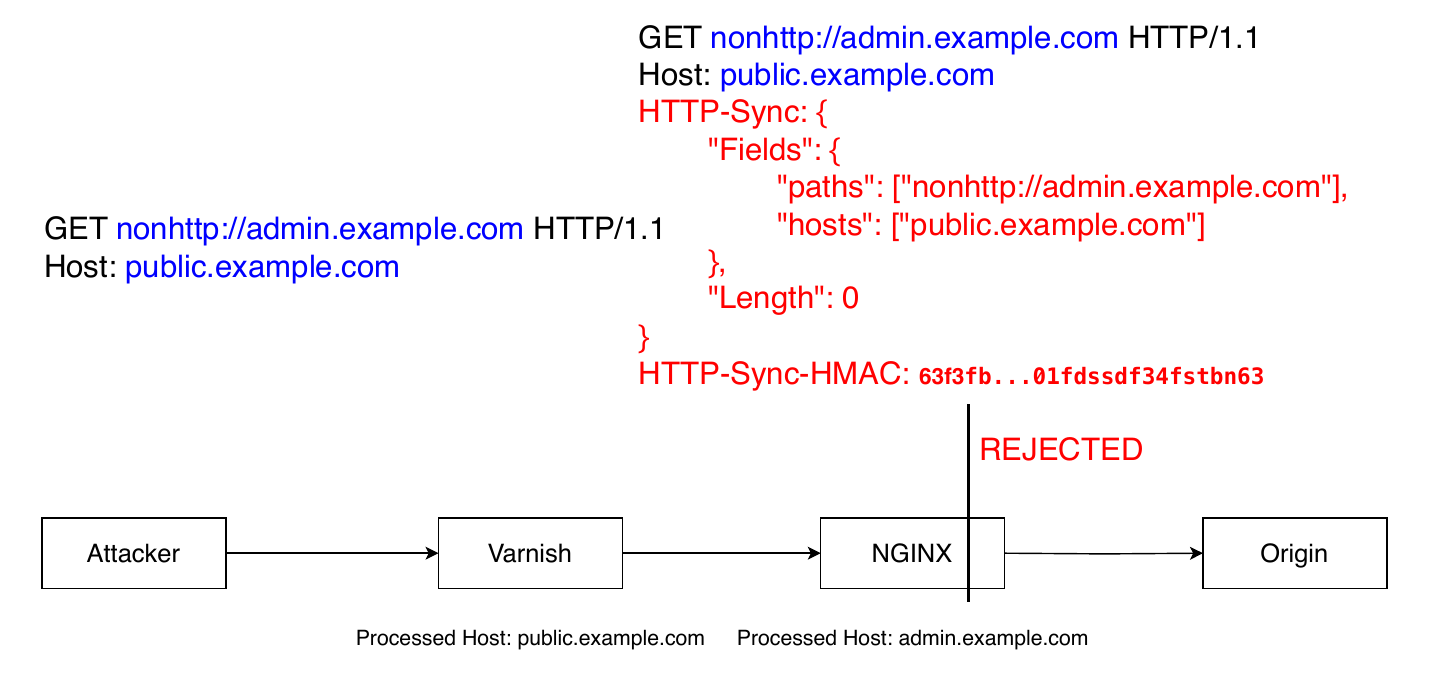}
\caption{Absolute URI Host Confusion Attack prevented.}
\label{fig:case_study_2}
\end{figure}

In this attack scenario, Varnish is the first hop in the traffic path tasked with preventing access to an administrative endpoint and NGINX serves as a second node. By abusing this host confusion, an attacker can bypass Varnish's security check and access the forbidden endpoint.
Figure~\ref{fig:case_study_2} demonstrates how our design prevents this attack. Once NGINX receives the request with the \defenseheader header, it first processes the request and determines the host name is \texttt{admin.example.com}. It then attempts to verify the host name that the previous hop (Varnish) sent, and spots the mismatch. NGINX then terminates the connection, thwarting the attack.

\subsubsection{Case Study 3: Request Framing Confusion}

In this case study, we implement and defend against an attack reported by Jabiyev et al.~\cite{jabiyev2021t}. This attack stems from a fat GET request (i.e., a GET request that also has a body) causing a request framing confusion issue. The vulnerable network configuration for this case study places NGINX as the first server in the traffic path, and Varnish as the second.

NGINX processes the fat GET request and includes the body in its forwarded request. Varnish will not consume the body; assuming it leaves the body in the connection buffer, for persistent connections, this discrepancy can lead to HRS.

\begin{figure}[t]
\centering
\includegraphics[width=0.79\linewidth]{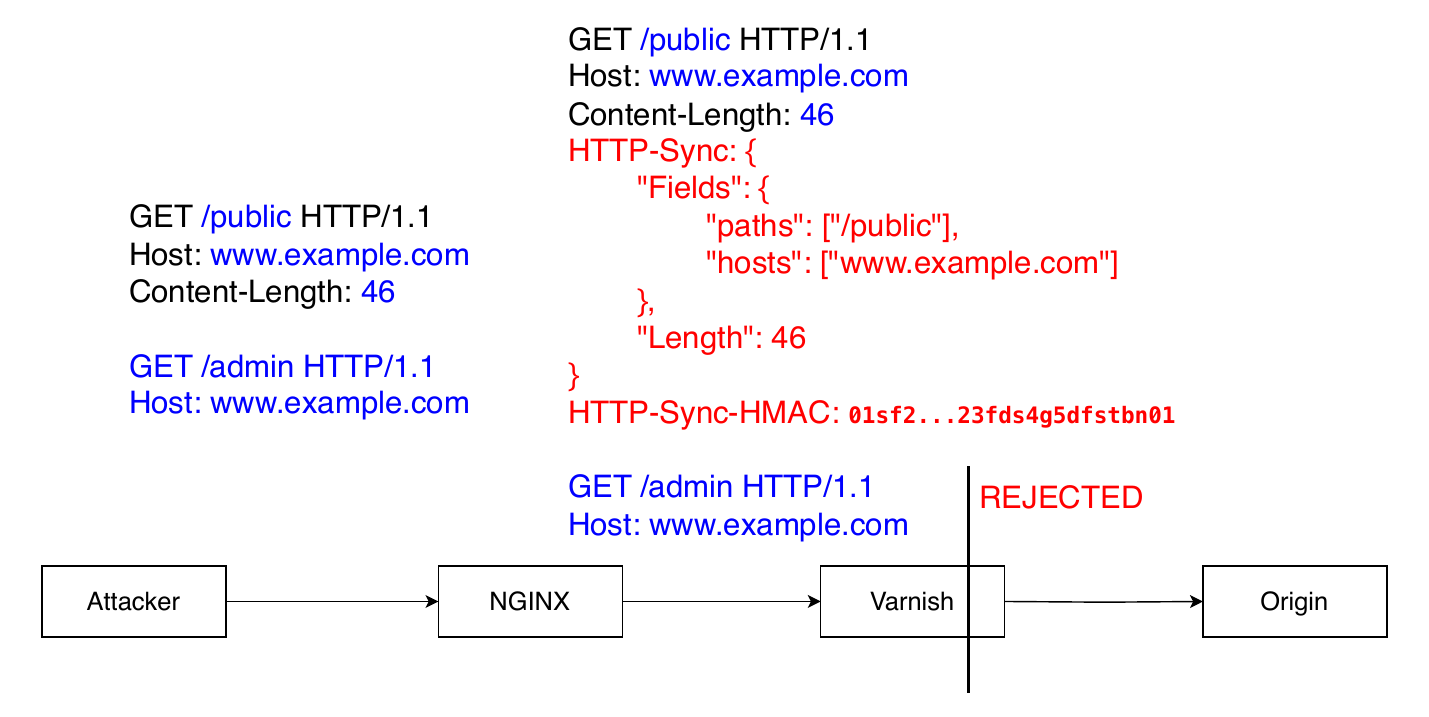}
\caption{Framing Confusion with a fat GET request prevented.}
\label{fig:case_study_3}
\end{figure}

Figure~\ref{fig:case_study_3} depicts how our design solves this problem. NGINX processes the request and fills the \defenseheader and \defenseheaderhmac headers. Varnish also processes the request, and observes that the processed request body length is 0. Varnish then compares the body length received in the \defenseheader with its processed length, and observes they are not equal, thus rejecting the request.

\section{Limitations}

HTTP Request Synchronization requires modifications to servers. However, we showed that these modifications can be easily implemented in both new and old versions of popular servers, demonstrating its feasibility with current and legacy systems.

Our system neither affects nor is affected by unaware servers in the communication path, while still providing protection for those that implement our methodology between the first and last implementing servers.

Key distribution is an essential limitation of our methodology that we cannot evade. However, we note that this is not a new problem that needs to be resolved specifically for our work. Existing distributed web architectures are likely to have deployed their own key management services for other secret sharing needs (e.g., Azure Key Vault~\cite{azure_key_vault}, AWS Key Management Service~\cite{aws_kms}, HashiCorp Vault~\cite{hashicorp_vault}). Hence, they can either adapt or introduce these systems for the necessary key distribution.

We use a standard web threat model, where all traffic is protected with TLS. In this setup, traffic manipulation and replay attacks are not feasible. The only attack that remains possible is proxy takeover, but the impact of such an attack is a complete collapse of the Internet security model, not just our defense.

We assume application-layer payloads remain unchanged (e.g., form fields are not altered). Our instrumentation could be extended for such cases by adding verification within modules. This is costly for us, but likely minor for their developers. Meanwhile, our system correctly handles transformations between Transfer-Encoding and Content-Length at any hop.

The communication path consists of servers configured by a single entity. This is not a limitation but reality -- third-party services like Cloudflare are fully configurable by you; otherwise, customer secrets passed through them would be exposed.
\section{Conclusion}
We have presented HTTP Request Synchronization, the first general defense to prevent HTTP processing discrepancies. Our defense scheme detects processing discrepancies resulting from path, host, and request framing confusion, i.e., all vulnerability vectors documented in literature so far, and can be extended in the future to increase its scope. We have presented a practical implementation of the approach for Apache, NGINX, HAProxy, Varnish, and Cloudflare, and demonstrated its practical value as a viable defense technique against an open security and safety problem.

\bibliographystyle{plain}
\bibliography{paper}

\appendix
\section*{Ethical Considerations}
This is a protocol and system design work proposing a defensive technique. All experiments and measurements were done in our lab environment. There are no ethical implications involving harm to any entity.

\section*{Open Science}
The server implementations with our defense strategy are located in \url{https://github.com/http-request-sync/defeats_discrepancy_attacks}.

\section*{CVEs Used For Security Evaluation}
\begin{table}[h]
\small
\centering
\caption{CVEs grouped by attack type.}
\label{tab:cves}
\begin{tabularx}{\linewidth}{lX}
\toprule
Attack Type & CVEs \\
\midrule
Path Confusion & 2025-52892, 2024-53008, 2024-27185, 2024-11234 \\
Host/Header Confusion & 2025-31137, 2025-41235, 2025-1736, 2025-0752, 2024-9666, 2024-56908, 2024-55925, 2024-29643, 2024-12397, 2025-0178, 2024-39736 \\
Framing Confusion & 2025-6442, 2025-53643, 2025-47905, 2025-4600, 2025-43859, 2025-32094, 2025-30346, 2025-22871, 2024-6827, 2024-53868, 2024-52530, 2024-52304, 2024-47220, 2024-35161, 2024-27982, 2024-23452, 2024-21647, 2024-1135, 2024-10264 \\
\bottomrule
\end{tabularx}
\end{table}

\begin{table}[h]
\centering
\caption{Server support for preserving trailer fields.
\label{tab:servers}}
    \begin{tabular}{lrrc}
    \toprule
        Server & Version & Preserves Trailers \\
        \midrule
        Apache httpd & 2.4.59  & No\\
        HAProxy & 3.0.2  & Yes\\
        NGINX & 1.27.0 & No\\
        Varnish & 7.5 & No\\
        \addlinespace
        Cloudflare & N/A & No\\
        \bottomrule
    \end{tabular}
\end{table}

\end{document}